\documentclass[conference]{IEEEtran}
\IEEEoverridecommandlockouts
\usepackage{cite}
\usepackage[
    colorlinks=true,
    linkcolor=blue,
    citecolor=blue,
    urlcolor=blue
]{hyperref}
\usepackage{amsmath,amssymb,amsfonts}
\usepackage[ruled,linesnumbered]{algorithm2e}
\usepackage{algorithmic}
\usepackage{graphicx}
\usepackage{textcomp}
\usepackage{ifthen}
\usepackage{xurl}
\usepackage{xcolor}
\usepackage{amssymb}
\newboolean{showcomments}
\setboolean{showcomments}{true} 
\def\BibTeX{{\rm B\kern-.05em{\sc i\kern-.025em b}\kern-.08em
    T\kern-.1667em\lower.7ex\hbox{E}\kern-.125emX}}

\makeatletter
\newcommand{\mynote}[3]{%
  \ifthenelse{\boolean{showcomments}}{%
   {\bfseries\sffamily\small#1:}
   {\small$\blacktriangleright$\textsf{\emph{\color{#3}{#2}}}$\blacktriangleleft$}}%
  {%
   \@bsphack
   \@esphack
  }%
}
\makeatother

\begin{document}

\title{AgileOS: A GPU Operating System Layer for Protected CUDA Services
\thanks{Note: This paper only presents our initial efforts towards a GPU OS. More design details, experimental results, and source code will be released upon the acceptance of the full version of this paper.}
}

\author{\IEEEauthorblockN{Zhuoping Yang}
\IEEEauthorblockA{\textit{Brown University} \\
\textit{Providence, Rhode Island, USA} \\
\textit{zhuoping\_yang@brown.edu} 
}
\and 
\IEEEauthorblockN{Yiyu Shi}
\IEEEauthorblockA{\textit{University of Notre Dame} \\
\textit{Notre Dame, Indiana, USA} \\
\textit{yshi4@nd.edu} 
}

\and 
\IEEEauthorblockN{Alex Jones}
\IEEEauthorblockA{\textit{Syracuse University} \\
\textit{Syracuse, New York, USA} \\
\textit{akj@syr.edu } 
}

\and 
\IEEEauthorblockN{Peipei Zhou}
\IEEEauthorblockA{\textit{Brown University} \\
\textit{Providence, Rhode Island, USA} \\
\textit{peipei\_zhou@brown.edu} 
}
}

\maketitle

\begin{abstract}
Modern GPU applications increasingly interact with storage systems, network devices, vendor libraries, and GPU-resident services rather than executing only isolated compute kernels.
This shift creates a need for operating-system-like protection around GPU services, where service metadata, device queues, memory-mapped I/O regions, and library-internal state should not be directly exposed to untrusted application kernels.
However, today's CUDA programming model, by default, still gives each application direct ownership of its CUDA context, device pointers, runtime handles, module loading path, and kernel launches, leaving protected GPU services to build their own ad hoc interfaces and isolation mechanisms.

This paper presents the initial design and prototype scope of AgileOS, a GPU operating-system layer for protected CUDA services.
AgileOS virtualizes CUDA at the library boundary: applications link against client-side CUDA Runtime, Driver, and selected library shims, while a trusted runtime worker owns the real CUDA context and mediates supported operations.
To protect service state and module interfaces, AgileOS also defines a GPU memory-management model that separates user allocations from protected module/MMIO ranges, using pointer validation and memory access guards via PTX injection.
AgileOS is modularized and flexible, supporting a range of protected services and existing libraries such as cuFFT and PyTorch.
The prototype includes client-side interceptors, worker-side CUDA handlers, virtualized CUDA object tables, protected AgileOS modules, a GPU memory manager that separates user allocations from protected module/MMIO ranges, selected trusted library adapters, and the PTX-level kernel memory guard.

\end{abstract}


\section{Introduction}

GPUs are increasingly used as general-purpose acceleration platforms in the AI era, driven by their programmability, massive parallelism, and rapidly advancing computational capability. While GPUs have been highly successful in deep learning~\cite{ye2024deep, menghani2023efficient, miao2025towards}, graph processing~\cite{liu2023bgl, yang2022graphblast, gong2023gsampler, chen2022efficient}, scientific computing~\cite{pandey2022transformational, xue2023jax, bayraktar2023cuquantum}, etc., modern GPU applications are becoming increasingly complicated and data-intensive. Rather than simply executing isolated compute kernels, these applications increasingly process large datasets that exceed local GPU memory capacity and are distributed across GPU clusters, making interaction with storage systems~\cite{qureshi2023gpu, chang2024gmt, yang2025agile, han2026asynchrony, qui2025geminifs, li2025gofs} and high-performance networks~\cite{wang2022fpganic, ren2025enabling} necessary.

Conventionally, storage and network access are mediated by the host CPU via an operating system (OS), such as Linux, which provides file systems, device drivers, virtual memory, permissions, process isolation, etc. Applications are allowed to use the underlying hardware resources via controlled system-call interfaces without directly managing the hardware state or trusting each other. 
While this OS boundary is effective and essential, the emerging GPU-centric systems have moved the third-party hardware management logic closer to the GPU, bypassing the CPU in the critical path. However, they do not inherit an equivalent GPU-side OS boundary, and thus, require the applications or ad hoc Interprocess Communication (IPC) mechanisms to own the control path, introducing security concerns or additional hardware resource usage.

This shift from compute-only acceleration to data- and system-facing execution creates a growing need for OS-like support on GPUs, including protected memory management, controlled I/O access, file-system integration, permission checks, and safe maintenance of GPU-resident runtime state.

Recent GPU OS efforts move beyond conventional GPU management, but they target different problems. GPUOS~\cite{yang2026gpuos} focuses on reducing kernel launch overhead by using a persistent GPU kernel, device-side task queues, and runtime operator dispatch to execute small operations without repeatedly crossing the host-device launch boundary. LithOS~\cite{coppock2025lithos} focuses on transparent GPU resource management for machine learning workloads, introducing fine-grained scheduling, kernel atomization, right-sizing, and power management to improve utilization, latency, and energy efficiency. These systems demonstrate that OS-like control layers can improve GPU execution efficiency and resource management. However, they do not directly address how CUDA applications should safely access protected GPU-resident system services such as file systems, I/O queues, memory managers, trusted library adapters, or permission modules. 

This paper proposes AgileOS, a GPU operating-system layer for protected CUDA services. AgileOS virtualizes CUDA at the library boundary, where applications link against client-side interceptor libraries that export CUDA Runtime, Driver, and selected vendor-library symbols, while actual GPU operations are forwarded to a trusted runtime. The runtime owns the real CUDA context through worker processes and becomes the mediation point for GPU memory allocation, module registration, kernel launches, stream and event operations, Driver API calls, and trusted library execution. This design relocates GPU context ownership from untrusted applications to a controlled OS-layer process, allowing AgileOS to validate API arguments, manage GPU resources, and expose GPU-resident services through a protected boundary. In this way, file systems, I/O queues, memory managers, trusted library adapters, and permission modules can be implemented as reusable AgileOS modules and services.

This paper makes the following contributions:
\begin{itemize}
    \item We present AgileOS, a GPU operating-system layer that virtualizes CUDA at the library boundary and moves real CUDA context ownership into trusted runtime workers.
    \item We design a protected CUDA service architecture with client-side interceptors, server-side API handlers, trusted library adapters, and configurable AgileOS modules.
    \item We introduce a GPU memory-management mechanism that separates AgileOS kernel/module memory and memory-mapped I/O (MMIO) from user allocations and enforces protection through pointer validation and optional PTX-level kernel memory guards.
    \item We implement a prototype supporting a practical subset of CUDA Runtime, Driver, and cuFFT functionality, including handled CUDA API names, isolated client containers without GPU device exposure, and regression tests covering CUDA forwarding, module calls, stream/event tokens, memory isolation, kernel guards, arrays, symbols, and cuFFT, etc.
\end{itemize}

\section{Background}

\subsection{GPU-Centric I/O Systems}
Modern GPU applications are no longer limited to computing over data that can be managed by GPU memory only. Large language models (LLMs), Large-scale graph analytics, recommendation models, scientific simulations, data analytics, and distributed AI workloads often operate on datasets that exceed local GPU memory capacity or require frequent communication with storage and network devices~\cite{gholami2024ai, rajbhandari2021zero, maurya2025mlp, zhang2023g10, yang2023overcoming, hu2025demystifying, wang2022fpganic, ren2025enabling}. In the conventional CPU-centric model, these interactions are mediated by the host operating system on the CPU. However, as the GPU becomes the dominant computation engine, keeping the CPU on the critical path for every storage or communication operation can introduce unnecessary latency and synchronization overhead. This has motivated a growing class of GPU-centric I/O systems that move portions of storage, memory-tiering, file-system, and network control closer to the GPU. 

Early GPU-initiated storage systems demonstrate the benefit of allowing GPU threads to participate directly in storage access. BaM~\cite{qureshi2023gpu} enables GPU threads to generate fine-grained storage requests and uses high-throughput queues and a GPU-side software cache to coalesce requests and reduce I/O amplification. GMT~\cite{chang2024gmt} extends this idea from direct storage access to GPU-orchestrated memory tiering, where the GPU coordinates data movement across GPU memory, host memory, and storage to support large working sets. AGILE~\cite{yang2025agile} further shows that GPU-SSD communication can be made asynchronous, allowing GPU computation and SSD I/O to overlap rather than forcing GPU threads to synchronously wait for storage completion. Similarly, AGIO~\cite{han2026asynchrony} identifies the mismatch between GPU execution and traditional synchronous I/O interfaces, and demonstrates when and why asynchronous I/O is useful on a comprehensive suite of dynamic applications. Together, these systems show that storage access is becoming part of the GPU execution model itself, rather than an external service invoked only by the CPU.  

Recent systems also move higher-level storage-management and file-system functionality closer to the GPU. GeminiFS~\cite{qui2025geminifs} provides a companion file-system interface for GPU programs, enabling direct file-based access to NVMe storage while coordinating metadata with the host file system. GoFS~\cite{li2025gofs} goes further by offloading scalable direct storage management to the GPU and exposing POSIX-like file-system APIs for GPU programs. This shift is significant because file systems and storage managers traditionally rely heavily on OS mechanisms such as access control, namespace management, metadata consistency, buffer ownership, and device-driver mediation. When similar functionality is moved into GPU-resident code or GPU-controlled runtime components, the system needs an equivalent protection and management boundary.  
  
A similar trend is visible in GPU-centric networking. FpgaNIC~\cite{wang2022fpganic} enables direct PCIe peer-to-peer communication between GPUs and an FPGA-based SmartNIC, allowing the GPU to interact with the network device using GPU virtual addresses. FuseLink~\cite{ren2025enabling} improves GPU communication over multiple NICs by integrating intra-server and inter-server communication paths and allowing GPU traffic to use otherwise idle NIC resources. These systems reflect the same broader architectural shift that GPUs increasingly interact with external devices and distributed-system resources directly, instead of treating the CPU as the sole orchestrator of I/O. As a result, GPU applications increasingly need mechanisms analogous to those provided by an operating system for CPUs, including controlled access to device queues, protected memory mappings, shared runtime state, and reusable service interfaces.  

\subsection{Existing GPU OSes and Abstractions}

Modern GPUs require more than a simple kernel-launch interface. As GPU workloads become increasingly latency-sensitive, multi-tenant, and service-oriented, both industry mechanisms and recent research systems have explored stronger GPU management abstractions. 

One direction of a GPU OS is reducing CPU-side kernel-launch overhead. In the conventional CUDA execution model, each kernel launch requires host-side runtime and driver work before the GPU can execute the kernel. This overhead is often acceptable for long-running kernels, but it becomes significant for workloads composed of many short GPU operations, such as fine-grained tensor operators, micro-batched inference, and parts of LLM serving. CUDA Graphs~\cite{nvidia_cuda_graphs} address this problem by capturing a sequence of GPU operations and replaying the captured graph as a single launchable unit, thereby amortizing repeated launch overhead across iterations~\cite{sglang_breakable_cuda_graph}. However, CUDA Graphs are most effective when GPU execution is repetitive and statically capturable, which limits their applicability to modern LLM inference. In the decode phase, each iteration often follows a similar per-token execution pattern, making graph capture effective for reducing kernel-launch overhead. In contrast, the prefill phase involves variable sequence lengths, changing tensor shapes, and dynamic batching or scheduling decisions, making it more difficult to capture and replay a single consistent graph. Systems such as Medusa~\cite{zeng2025medusa} further show that CUDA Graphs introduce important initialization and materialization challenges in serverless LLM serving, where graph capture and restoration become part of the cold-start path. GPUOS~\cite{yang2026gpuos} explores a different solution to the same broad launch-overhead problem. Instead of repeatedly launching many small kernels, GPUOS uses a persistent GPU kernel, device-side task queues, and runtime operator injection to execute small operations inside a long-running GPU-resident execution substrate. This design demonstrates that OS-like GPU execution control can reduce launch overhead and improve efficiency for fine-grained operations.

The second direction of a GPU OS focuses on transparent GPU sharing and multi-tenancy. NVIDIA MPS~\cite{nvidia_mps} allows multiple CUDA processes to cooperatively share a GPU through a runtime service, improving utilization and reducing context-switching overhead among CUDA applications. NVIDIA MIG~\cite{nvidia_mig} provides stronger partitioning by dividing supported GPUs into multiple isolated instances with dedicated compute and memory resources. These mechanisms are useful for production GPU clusters, but they provide limited flexibility. 
MPS improves cooperative sharing but does not provide strong per-application protection for arbitrary untrusted workloads, while MIG provides hardware isolation but relies on relatively coarse and fixed resource partitions. 
Research systems therefore explore more flexible sharing mechanisms. TGS~\cite{wu2023transparent} provides transparent GPU sharing beneath containers through user-space interception and combines adaptive rate control with transparent unified memory to improve utilization while preserving performance isolation. While runtime- or user-space interception has been applied to many GPU systems~\cite{wu2023transparent, wang2024characterizing, gu2024gvulkan, chow2023krisp}, KRYPTON~\cite{zhang2025efficient} avoids the substantial engineering effort required to track diverse runtime APIs by intercepting GPU command buffers in kernel space, thereby providing virtual GPU devices with spatio-temporal sharing. Together, these systems illustrate a shift toward lower-layer GPU resource management to improve transparency, compatibility, and control. Extending this perspective, LithOS~\cite{coppock2025lithos} advocates treating GPU management as an operating-system responsibility for machine-learning workloads. It introduces fine-grained TPC scheduling, kernel atomization, hardware right-sizing, and power management to improve utilization, latency, and energy efficiency in multi-tenant ML environments.

However, existing GPU OSes and sharing directions still leave an important abstraction gap. They primarily manage how GPU computation is launched, scheduled, partitioned, or multiplexed among applications. Unlike traditional operating systems such as Linux, which enforce a clear separation between user space and kernel space through protected interfaces, these systems do not directly address how CUDA applications should safely access protected GPU-resident system services. This gap becomes increasingly important for GPU-centric storage, file-system, and networking systems, where GPU-resident components may need to maintain private metadata, protected memory regions, device queues, permission state, and trusted library handles.

\subsection{CUDA Memory Protection and Limitations}

Modern GPUs support virtual memory, MMUs, TLBs, and page tables, but the protection abstraction exposed to CUDA applications is still different from a traditional CPU operating-system model. On CPUs, the OS relies on privilege levels, page permissions, system-call boundaries, and user/kernel address-space separation to protect kernel memory and device-driver state from applications. In contrast, CUDA exposes the GPU through application-owned contexts. An application allocates device memory, receives raw GPU pointers, creates streams and events, loads modules, and launches kernels within its own context. Although GPU virtual memory provides address translation and context-level isolation, it does not provide a standard user/kernel split inside a CUDA context for protecting GPU-resident services.

This limitation matters for GPU-centric system services. A GPU-side file system, I/O queue manager, memory manager, or trusted library adapter may need to maintain private metadata, protected buffers, and memory-mapped device state. However, if these regions are placed in the same CUDA-accessible execution environment as untrusted application kernels, CUDA does not naturally prevent those kernels from attempting to access or corrupt them. Prior work on GPU memory exploitation further shows that modern GPUs lack several CPU-style memory-protection mechanisms and that memory errors can lead to cross-space corruption, code injection, and code-reuse attacks~\cite{guo2024gpu}. Therefore, protected GPU services require an additional software-defined boundary beyond conventional CUDA memory management.
\section{Threat and Trust Model}

AgileOS assumes that client applications run in containers without direct access to physical GPU devices, third-party PCIe devices, or the host CUDA installation.
All supported GPU-facing operations therefore enter through AgileOS CUDA-compatible libraries and are forwarded to a trusted AgileOS runtime.
The host operating system, NVIDIA driver, custom AgileOS drivers, runtime daemon, scheduler, workers, trusted adapters, and module services are trusted.
Client applications, user kernels, and client containers are untrusted.

An untrusted client may issue arbitrary supported CUDA Runtime, Driver, module-control, or selected library calls; pass invalid, stale, protected, or cross-client pointers; forge virtualized handles; invoke libraries with unauthorized buffers; or launch kernels that compute addresses inside AgileOS-protected ranges.
AgileOS aims to prevent clients from bypassing the runtime, directly accessing protected GPU service state, corrupting AgileOS kernel/module memory, using CUDA objects owned by other clients or services, or exposing protected MMIO and service metadata through ordinary CUDA APIs.

AgileOS does not defend against a compromised host kernel, compromised GPU driver, malicious administrator, physical attacks, or all GPU microarchitectural side channels. 
The current prototype does not claim complete CUDA API coverage, full cuFFT/PyTorch compatibility, a complete GPU file system, or performance isolation, and we will treat them as our active ongoing research directions. 
\section{AgileOS Design}



\subsection{Design Goals and Principles}
AgileOS has three primary design goals. First, it should provide a protected boundary between untrusted CUDA applications and trusted GPU-resident services. Trusted services may manage persistent metadata, memory-mapped I/O regions, device queues, or library-internal state, and applications should only access this state through controlled AgileOS interfaces. Second, AgileOS should preserve CUDA compatibility as much as possible. Applications should not need to be rewritten around a new programming model; instead, AgileOS should mediate existing CUDA Runtime, Driver, and library APIs through client-side compatibility layers. Third, AgileOS should support modular service composition, allowing applications to use only the protected services they require while keeping the trusted computing base small.

These goals lead to four design principles. First, \emph{context ownership must be trusted}: the real CUDA context is owned by AgileOS rather than by the application. Second, \emph{CUDA operations are mediated requests}: CUDA calls are treated as controlled entries into AgileOS, analogous to system calls in a traditional operating system. Third, \emph{CUDA objects are virtualized}: applications receive opaque handles rather than directly manipulating server-side CUDA objects. Fourth, \emph{service memory is protected}: AgileOS separates user allocations from AgileOS kernel memory, module-private memory, and MMIO regions, and checks accesses through software mediation.




\subsection{System Overview}

\begin{figure}
    \centering
    \includegraphics[width=1\linewidth]{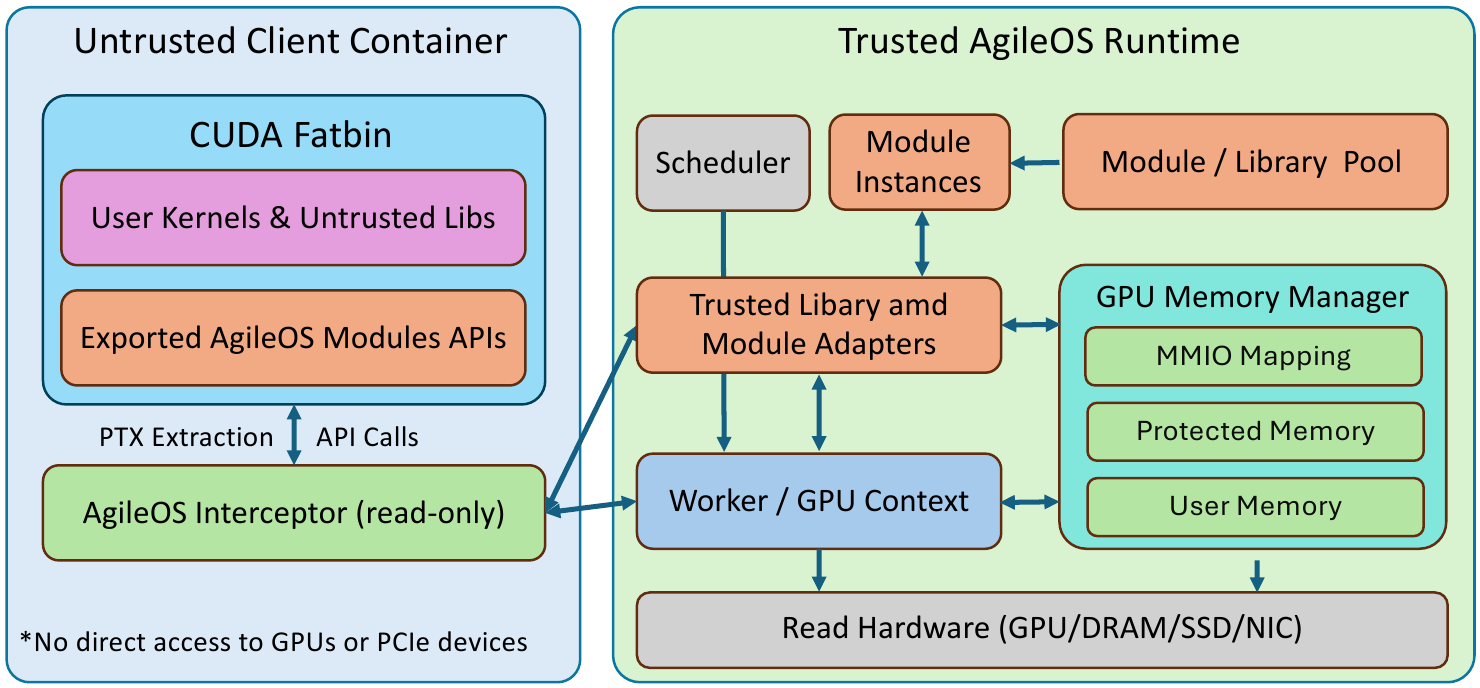}
    \caption{AgileOS system overview. Untrusted CUDA applications use the AgileOS CUDA interceptor to access a trusted runtime that owns the real CUDA context, mediates CUDA operations, manages protected GPU memory, and exposes protected module services.}
    \label{fig:system-overview}
\end{figure}

Figure~\ref{fig:system-overview} shows the high-level architecture of AgileOS.
AgileOS separates an untrusted client container from a trusted AgileOS runtime.
The client container runs the CUDA application, including user kernels, untrusted libraries, and the exported AgileOS module APIs. The container does not have direct access to GPUs or PCIe devices. Instead, all CUDA-facing operations enter AgileOS through a read-only client-side interceptor, which preserves the CUDA programming interface while forwarding requests and the PTX source code from the user-compiled CUDA fat binaries to the trusted runtime.

The trusted AgileOS runtime owns the real GPU context and mediates all accesses to hardware resources. A scheduler assigns each client to a worker, and the worker executes CUDA operations in its owned GPU context. The worker interacts with trusted library and module adapters, which provide controlled server-side execution for selected vendor libraries and AgileOS modules. Module instances are created from a configured module pool and hold per-client trusted state. This design ensures that applications can invoke public module APIs without receiving direct access to protected module state, service sockets, or device resources.

AgileOS also centralizes GPU memory management (Section~\ref{sec:memory_protection}) inside the trusted runtime. The GPU memory manager separates user memory from protected AgileOS memory and memory-mapped I/O regions. User allocations are exposed to the application through virtualized CUDA pointers, while protected memory and MMIO mappings are used only by trusted workers, module instances, and adapters. The worker and memory manager jointly validate pointer arguments, translate virtualized handles, and enforce that untrusted kernels access protected resources only through approved AgileOS module interfaces. 

\subsection{CUDA Virtualization Boundary}
AgileOS virtualizes CUDA at the library boundary.
Instead of linking directly against vendor CUDA libraries, applications link against or preload AgileOS-provided CUDA-compatible shims.
Each supported CUDA Runtime, CUDA Driver, module-control, or trusted-library operation is serialized as a request to the worker.
Unsupported operations should fail closed.

On the server side, the worker dispatches requests to CUDA handlers that operate on the real CUDA context.
Client-visible CUDA objects are tokens backed by worker-owned tables, which allows AgileOS to reject fabricated, stale, or cross-client handles.
User allocations are created by the worker through a GPU memory manager enforcing a protected GPU memory model (Section~\ref{sec:memory_protection}).
When a client later passes a pointer to a CUDA operation or trusted adapter, the worker can determine whether the pointer belongs to a live user allocation, an AgileOS-protected range, or an unknown address. Similar to other user-space interception systems~\cite{wu2023transparent, wang2024characterizing, gu2024gvulkan, chow2023krisp}, this mediation requires huge engineering efforts, and we avoid it via an agentic implementation and test flow (Section~\ref{sec:agentic_flow}).

Kernel and module loading follow the same principle.
Application fat binaries or PTX arrive at the worker before code is loaded into the real CUDA context.
The worker may replace public AgileOS module placeholders with trusted implementations from configured module artifacts.
When enabled, the optional kernel memory guard instruments application PTX so direct accesses to AgileOS-protected ranges trap.

\subsection{Module and Service Model}
\label{sec:module_service}
AgileOS modules are the protected-service abstraction exposed to applications.
A module contains a public application-facing interface with a trusted AgileOS plugin implementation and, optionally, a privileged service process.
This separation allows applications to call GPU-resident services without receiving direct access to service metadata, device queues, MMIO mappings, or module-private state.

\noindent \textbf{Client-scoped selection:} 
Module use is declared at the client boundary rather than granted implicitly.
An application exposes its requested public module interfaces to AgileOS by linking with public module header files, and the worker admits them only if they match the deployment's configured module policy, version requirements, and dependency constraints.
As a result, module state is created only for modules that are both requested by the client and approved by the worker.
This keeps service composition explicit and bounds the trusted code and protected state associated with each client.

\noindent \textbf{Protected module state:}
For each admitted module, AgileOS creates a per-client trusted module instance in protected GPU memory.
The application sees only the public interface; it does not receive the module object's address, service IPC interface, shared-memory handle, physical address, or MMIO pointer.
Module-private state can therefore hold sensitive service metadata.
The protected memory model in Section~\ref{sec:memory_protection} is responsible for separating these module ranges from ordinary user allocations.

\noindent \textbf{GPU-side invocation:}
Modules may export public device functions that application kernels can call.
The application binary contains the public call sites, while AgileOS binds those calls to trusted module implementations during mediated module loading at the PTX level.
The trusted implementation can access the module's protected state, but the calling application kernel cannot directly discover or dereference that state (Section~\ref{sec:memory_protection}).
Thus, a user kernel can request a module operation through an approved function while direct access to protected module memory remains outside the application's CUDA addressability contract.

\noindent \textbf{Host-side control:}
Modules may also expose host-side control commands for configuration and resource acquisition.
These commands use copied request and response payloads rather than pointers into the application process, which allows the worker to validate the selected module, command name, payload size, and service policy before invoking trusted module code.
For example, a storage module may accept a public ``select device'' command and return public queue metadata, while the actual queue lease, device mapping, and service-owned shared region remain hidden in AgileOS-protected state.

\noindent \textbf{Service-backed modules:}
Some modules require resources that should persist beyond a single CUDA call or require privileges not granted to the client container, required in asynchronous GPU-centric storage access system for decoupling the SQ and CQ management~\cite{yang2025agile, han2026asynchrony}.
AgileOS supports these cases with module services: trusted helper processes launched from the AgileOS configuration.
A service can maintain persistent device or controller state and allocate, lease, or share resources among per-client module instances.
The service is not exposed to the application; all client-visible operations still pass through the worker and the selected module instance.

\noindent \textbf{Lifetime and revocation:}
Module state is scoped to the client-worker assignment and is released when the assignment ends.
Resources granted to that client, such as queue leases, shared GPU regions, or device attachments, are revoked as part of module teardown.
The underlying service remains an AgileOS-managed component and can continue serving other clients after the revoked client state is removed.
If a client violates the protected-memory policy and the worker context can no longer be trusted, AgileOS can retire that worker while still revoking the resources associated with the client's module instances.

\subsection{Protected GPU Memory Model}
\label{sec:memory_protection}
\begin{table*}[t]
\centering
\footnotesize
\caption{Prototype AgileOS GPU virtual-address layout.}
\label{tab:memory-layout}
\begin{tabular}{|p{0.28\textwidth}|p{0.1\textwidth}|p{0.52\textwidth}|}
\hline
Virtual Memory Address Range & Size & Owner and Usage \\
\hline
\texttt{0xF0000000000}--\texttt{0xF1000000000} & 64 GiB & Protected I/O aperture for MMIO mappings \\
\hline
\texttt{0xF1000000000}--\texttt{0xF8000000000} & 448 GiB & Protected AgileOS kernel/module memory and imported service memory \\
\hline
\texttt{0xF8000000000}--\texttt{0x108000000000} & 1 TiB & User allocations returned through CUDA-facing APIs \\
\hline
\end{tabular}
\end{table*}

AgileOS separates GPU memory into regions with different trust and ownership semantics. Table~\ref{tab:memory-layout} shows the prototype virtual memory layout.
User allocations requested through CUDA-facing APIs are placed in an AgileOS-managed user region and returned to applications as ordinary CUDA device pointers.
Protected AgileOS memory stores module objects, module-private metadata, imported service-owned shared GPU regions, and runtime state.
A protected I/O aperture is used for MMIO mappings such as PCIe BARs registered for GPU access.

This layout gives AgileOS a simple classification rule for mediated CUDA operations.
When a client passes a device pointer to a CUDA handler or trusted library adapter, the worker resolves the pointer against live user allocations and verifies that the requested byte range remains within that allocation.
Operations on valid user ranges can proceed, while operations that point to protected module memory, imported service queues, MMIO mappings, freed allocations, or unknown addresses fail before reaching the real CUDA or vendor-library call.
This check is especially important because protected and user regions live in the same worker-owned CUDA context, and the boundary is enforced by AgileOS mediation rather than by a hardware user/kernel privilege bit inside the CUDA context.

For service-backed modules, the protected ranges also serve as the attachment point for resources created outside the application.
A service may create shared GPU memory or provide an opened device resource, and the trusted module may store the resulting GPU addresses, queue identifiers, or doorbell pointers in its private state.
The application receives only public command results or ordinary user pointers; it never receives the exported handle, service socket, physical address, BAR address, or MMIO pointer.

Kernel launches require an additional guard because an untrusted kernel can construct arbitrary addresses once it is running. AgileOS instruments application PTX with a memory guard that traps direct global-memory accesses into the protected I/O and kernel/module ranges (Section~\ref{sec:memory_guard}). Thus, AgileOS combines API-time pointer validation with kernel-time access checks to protect GPU-resident service state on both the host CPU and the GPU side.

\subsection{Trusted Library}

Trusted library adapters preserve familiar vendor-library APIs while moving selected library execution from the client process into the trusted worker.
Applications continue to call library APIs through AgileOS shims, but real library state and real library handles remain server-side.
This complements the module model in Section~\ref{sec:module_service}: modules expose new protected services, while trusted library adapters provide compatibility for existing vendor-library interfaces.

The worker admits only configured adapters and configured library operations, so an untrusted application cannot choose arbitrary worker-side code or request arbitrary vendor-library entry points.
Before a trusted adapter invokes a vendor library, AgileOS checks that all handles and pointer arguments refer to objects owned by the current client and to live user allocations rather than protected AgileOS memory.
This turns a library call into a mediated operation with the same protection principles as CUDA calls. Similarly, the mediation layer can be implemented and tested via the agentic flow described in Section~\ref{sec:agentic_flow} to ease the burden on human efforts.

\section{Implementation of AgileOS}

\subsection{CUDA Virtual Memory Management}
\label{sec:vmm}

\begin{figure*}[tbh]
    \centering
    \includegraphics[width=1\linewidth]{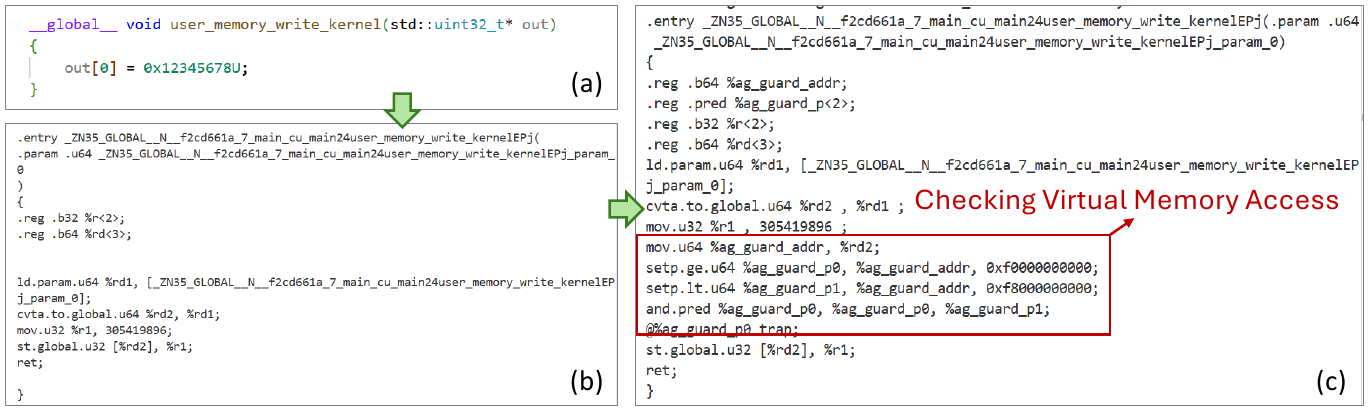}
    \caption{PTX-level kernel memory guard; (a) a CUDA kernel in the user program; (b) received PTX code from the compiled user program in AgileOS; (c) instruments the PTX with an address-range check before accessing the global memory, preventing direct accesses into the AgileOS protected I/O and kernel/module ranges while leaving ordinary user allocation accesses on the normal path.}
    \label{fig:ptx-mem-guard}
\end{figure*}

AgileOS relies on CUDA virtual memory management (VMM) APIs~\cite{nvidia_cuda_vmm} to construct the virtual-address layout in Table~\ref{tab:memory-layout}.
At worker initialization, the GPU memory manager first decide the granularity by checking the device allocation granularity using \texttt{cuMemGetAllocationGranularity}.
It then reserves two fixed virtual-address ranges with \texttt{cuMemAddressReserve}, including one range for AgileOS protected kernel/module memory and another for user allocations returned through CUDA-facing APIs.

Physical GPU memory is attached to those reserved ranges lazily.
For user allocations, AgileOS creates a CUDA allocation handle with \texttt{cuMemCreate}, maps it into the user aperture with \texttt{cuMemMap}, and grants device access with \texttt{cuMemSetAccess}.
The worker records the returned base address and requested size in an allocation table.
For protected module state, AgileOS uses the same VMM machinery but maps memory from the protected kernel/module aperture.
The worker also keeps host-side mappings for selected protected allocations so trusted host module code can initialize module objects and metadata.

Imported service memory uses a similar path.
When a module service creates a shareable GPU allocation, the worker imports the service-provided file descriptor with \texttt{cuMemImportFromShareableHandle}, maps the imported handle into an unused protected kernel/module subrange, and updates the module's protected state with the resulting GPU address.
The application never receives the shared-memory file descriptor or the protected GPU address.

\subsection{Controlled MMIO Mapping}
MMIO mappings require a different path because they originate from host device files rather than CUDA allocation handles.
For service-backed modules such as NVMe, the trusted module asks AgileOS to map an opened device resource into the protected I/O aperture.
The memory manager chooses the next page-aligned virtual address inside the I/O aperture and passes that address as the suggested target to \texttt{mmap}.
The implementation uses a fixed-address mapping policy, currently through \texttt{MAP\_FIXED\_NOREPLACE}, so the kernel either maps the region at the requested address or fails.
After \texttt{mmap} returns, AgileOS explicitly checks that the returned CPU pointer is exactly the requested I/O-aperture address; if not, it unmaps the result and rejects the mapping.

Once the CPU mapping is established, AgileOS registers it for GPU access with \texttt{cudaHostRegister} using I/O-memory registration flags, then obtains the GPU-visible pointer with \texttt{cudaHostGetDevicePointer}.
The CPU address, GPU address, size, and debug name are tracked in the worker.
Only trusted module code receives this metadata.
This check-and-record discipline prevents accidental placement of an MMIO mapping outside the protected aperture and gives the worker enough state to unregister and unmap the region during cleanup.

\subsection{Pointer Metadata and DMA Registration}
Every user allocation is tracked by base address and live size.
CUDA memory-copy handlers, Driver API handlers, and trusted library adapters resolve a client-supplied device pointer by searching this table and checking that the requested byte range remains inside one live allocation.
Pointers into the I/O aperture, protected module memory, freed allocations, or unknown addresses are rejected before the worker calls the real CUDA API or a vendor library.

The same metadata is used for DMA registration.
A service-backed module may ask AgileOS to register a user buffer for device DMA, but the request is accepted only for an exact live user allocation.
AgileOS pins the backing GPU allocation through its kernel driver, records a registration identifier, and returns the physical DMA information only to trusted module state.


\subsection{Module and Service Implementation}

AgileOS loads modules as configured trusted components, with an optional companion service when a module needs long-lived host-side state or access to external devices.
During client registration, the worker resolves the module dependency set, reads the selected module PTX, and creates per-client module objects in protected GPU memory.
For the host portion of a module, the worker opens the configured \texttt{lib<name>.so} with \texttt{dlopen}, resolves the exported \texttt{host\_init\_module} entry with \texttt{dlsym}, and passes an initialization context containing allocation and service callbacks.
The returned GPU address is written into the protected device-visible module table, while the worker keeps the CPU mapping needed for later host control and cleanup.

Host-side module commands use the same dynamic Application Binary Interface (ABI).
When an application calls the public module-control path, the worker verifies that the module is configured and selected, checks the copied payload size, resolves \texttt{host\_module\_control}, and invokes it with the already-created module object.
During teardown, the worker resolves \texttt{host\_cleanup\_module} and invokes modules in reverse dependency order so service leases, DMA registrations, and protected allocations can be released.
If the module is backed by a service, these callbacks acquire or release service-owned resources and install only private metadata into protected module state.

\subsection{PTX Processing and Kernel Memory Guard}
\label{sec:memory_guard}

AgileOS handles application PTX at the CUDA module-registration boundary.
When the interceptor forwards a Runtime fatbin registration, the worker records the PTX payload and delays the actual CUDA load until the matching registration-end call.
The worker then parses the PTX, applies AgileOS transformations, serializes the result, and loads the transformed module into the worker-owned CUDA context.
Selected module PTX is prepared during module registration, so application placeholders emitted by public module headers can be replaced with trusted function bodies when the application PTX is loaded.

The kernel memory guard is applied after trusted-function injection and before CUDA loads the transformed PTX.
When enabled, the guard instruments application global-memory accesses against the protected I/O and kernel/module intervals, while skipping registered trusted module functions. An example of \texttt{trap} memory gaurd is shown in Figure~\ref{fig:ptx-mem-guard}.
AgileOS supports different guard policies as follows:
\begin{itemize}
    \item \textbf{\texttt{trap}.} Trap immediately on protected-memory access; this is the safest default, but the worker context may need to be retired.
    \item \textbf{\texttt{remap\_or}.} Rewrite protected addresses into the user range for compatibility experiments; unsafe unless the remapped target is also validated as a live user allocation.
    \item \textbf{\texttt{poison}.} Record a violation and fail at a later synchronization or launch boundary when possible.
    \item \textbf{\texttt{sink}.} Redirect illegal accesses to quarantine memory for debugging without preserving isolation guarantees.
    \item \textbf{\texttt{report}.} Count or log would-be violations without enforcing, useful for auditing coverage but not isolation-safe.
\end{itemize}

\subsection{Agentic API Intercepting and Testing Flow}
\label{sec:agentic_flow}
AgileOS forwards CUDA Runtime, Driver, and selected library calls through client-side shims and worker-side handlers.
This boundary is intentionally strict, and client code must not fall back to host CUDA libraries, because it would bypass pointer checks, handle virtualization, and trusted-library mediation.
The challenge is that CUDA compatibility requires many individual hooks.
While some calls are simple queries, many of them need AgileOS-specific state for allocation, pointer validation, module registration, kernel launch, streams/events, arrays, symbols, or trusted library handles.

We therefore implement API support with the two-stage agentic flow shown in Algorithm~\ref{alg:agentic-testing-flow}.
Stage 1 is an LLM-agent iteration loop over a bounded API batch.
For each API, the agent writes a forwarding contract that records argument directions, pointer ownership, expected CUDA errors, required worker state, and memory-safety invariants.
It then generates focused tests, edits the category-scoped interceptor and worker handlers, and retries until the generated tests pass without violating mediation invariants.
The accepted batch updates the agent memory with supported APIs and checked invariants.

Stage 2 moves from generated tests to human-in-the-loop validation on real CUDA programs.
CUDA Samples and other high-quality human-written programs run inside a constrained Docker container that exposes the AgileOS interceptor but no direct CUDA path.
All CUDA calls must therefore pass through the worker.
Failures from this stage are manually reviewed; checker agents may summarize traces and coverage, but they are not allowed to patch the implementation.

This separation keeps the two validation goals distinct.
Stage 1 is optimized for fast implementation and invariant-focused regression tests, while Stage 2 checks whether the resulting compatibility surface works for realistic applications.
Each batch produces an overlay manifest listing the changed files, making the generated implementation easy to review, merge, or discard.

\begin{algorithm}[t]
\footnotesize
\DontPrintSemicolon
\SetKwProg{Fn}{function}{}{end function}
\SetKwFunction{LLMStage}{Stage1\_LLMTesting}
\SetKwFunction{HumanStage}{Stage2\_HumanValidation}
\SetKwInOut{Input}{Input}
\Input{API batch, human scaffold, agent memory, program corpus}
\Fn{\LLMStage{API batch}}{
    \ForEach{API in the batch}{
        write forwarding contract for API\;
        \Repeat{tests pass and invariants hold}{
            generate focused tests from the contract\;
            implement or revise interceptor and worker handlers\;
            run generated tests\;
            \eIf{tests fail}{
                summarize failure in agent memory\;
            }{
                \eIf{patch violates mediation invariants}{
                    reject patch\;
                }{
                    record API and checked invariants\;
                }
            }
        }
    }
}
\BlankLine
\Fn{\HumanStage{program corpus}}{
    \ForEach{program in the corpus}{
        run program in constrained AgileOS container\;
        \If{program fails}{
            summarize trace and coverage\;
            \eIf{manual review finds unsupported API}{
                enqueue API for next batch\;
            }{
                record compatibility or environment issue\;
            }
        }
    }
}
\caption{Two-stage agentic testing flow for CUDA API forwarding}
\label{alg:agentic-testing-flow}
\end{algorithm}
\subsection{Prototype Scope}
The AgileOS prototype now implements the core execution path needed for protected CUDA services: CUDA Runtime and Driver forwarding, module-control paths, service-backed module examples, a cuFFT trusted-library adapter, and the PTX-level kernel memory guard.
We are actively expanding API coverage through the agentic forwarding flow and validating the system with CUDA Samples, targeted regression tests, and higher-level application code.
Although full CUDA and vendor-library compatibility remains ongoing work, the current implementation already demonstrates that CUDA-facing compatibility, controlled virtual memory layout, service-backed protected modules, and trusted library adapters can coexist in one worker-owned CUDA context without exposing protected GPU service state to the client.

\section{Related Works}

\noindent \textbf{GPU-side storage and I/O.}
Recent storage systems reduce CPU involvement by allowing GPU execution to interact more directly with storage and memory-tiering mechanisms.
BaM~\cite{qureshi2023gpu} enables GPU-initiated storage access, GMT~\cite{chang2024gmt} uses GPUs to orchestrate memory tiering, and AGILE~\cite{yang2025agile} and AGIO~\cite{han2026asynchrony} explore asynchronous GPU--SSD interaction.
GeminiFS~\cite{qui2025geminifs} and GoFS~\cite{li2025gofs} further expose file-system or storage-management abstractions to GPU programs.
A similar direction appears in GPU-centric networking, where FpgaNIC~\cite{wang2022fpganic} enables GPU communication with an FPGA-based SmartNIC, and FuseLink~\cite{ren2025enabling} improves GPU communication over multiple NICs.
These systems motivate the need for GPU-side services, but they are primarily concerned with the design and performance of the I/O path.
AgileOS is complementary and focus on how such services can be exposed to untrusted CUDA applications through a protected worker context, mediated CUDA APIs, module-private state, and trusted library adapters.

\noindent \textbf{Accelerator-side system services.}
The shift toward near-device system services is broader than GPUs.
SmartNIC and multi-NIC GPU communication systems show that accelerators and attached devices can participate directly in networking and data movement~\cite{wang2022fpganic,ren2025enabling}.
FPGA-based storage systems provide another representative line of work beyond the GPU domain.
DONGLE~\cite{wong2023dongle,wong2024dongle} use FPGAs to orchestrate direct NVMe storage access for high-level synthesis workloads, while HiLFS~\cite{na2026hilfs} extends this direction with FPGA-orchestrated file-system services.
Together, these systems point to a broader trend in which system functionality moves from the CPU into accelerator-adjacent runtimes or devices.
AgileOS follows this trend for CUDA applications, but focuses specifically on the protection problem that arises when accelerator-side services must maintain private GPU memory, MMIO mappings, queue state, and library metadata in the presence of untrusted GPU kernels.
\bibliographystyle{IEEEtran}
\bibliography{bib/reference}
\end{document}